# RELATIVE SQUARED DISTANCES TO A CONIC

Valere Huypens, Belgium

## Abstract

*The midpoint method or technique is a "measurement" and as each measurement it has a tolerance, but worst of all it can be invalid, called Out-of-Control or OoC. The core of all midpoint methods is the accurate measurement of the difference of the squared distances of two points to the "polar" of their midpoint with respect to the conic. When this measurement is valid, it also measures the difference of the squared distances of these points to the conic, although it may be inaccurate, called Out-of-Accuracy or OoA. The primary condition is the necessary and sufficient condition that a measurement is valid. It is comletely new and it can be checked ultra fast and before the actual measurement starts.*

*Modeling an incremental algorithm, shows that the curve must be subdivided into "piecewise monotonic" sections, the start point must be optimal, and it explains that the 2D-incremental method can find, locally, the global Least Square Distance. Locally means that there are at most three candidate points for a given monotonic direction; therefore the 2D-midpoint method has, locally, at most three measurements.*

*When all the possible measurements are invalid, the midpoint method cannot be applied, and in that case the ultra fast "OoC-rule" selects the candidate point. This guarantees, for the first time, a 100% stable, ultra-fast, berserkless midpoint algorithm, which can be easily transformed to hardware. The new algorithm is on average (26.5±5)% faster than Mathematica, using the same resolution and tested using 42 different conics. Both programs are completely written in Mathematica and only ContourPlot[] has been replaced with a module to generate the grid-points, drawn with Mathematica's Graphics[Line{gridpoints}] function.*

## 1. POINT LATTICE — DIRECTED POLAR — PROPERTIES OF CONICS (FIG.1., FIG.2.)

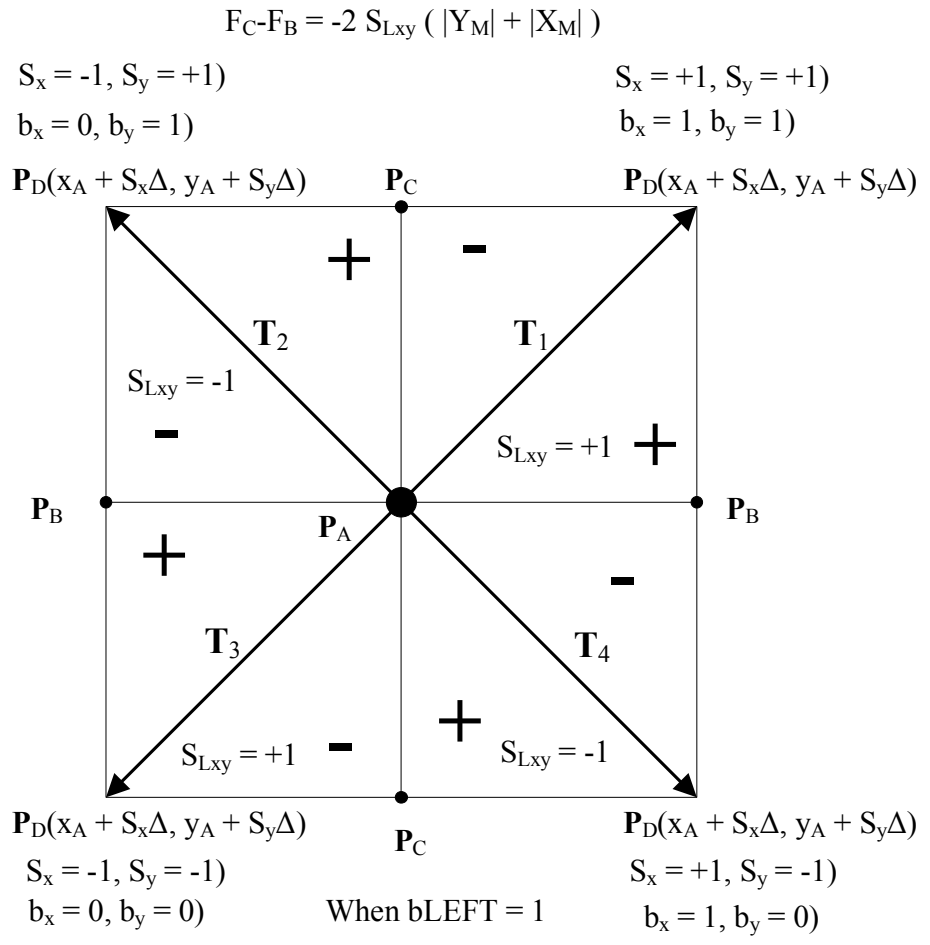

Fig. 1. The monotonic vectors Ti

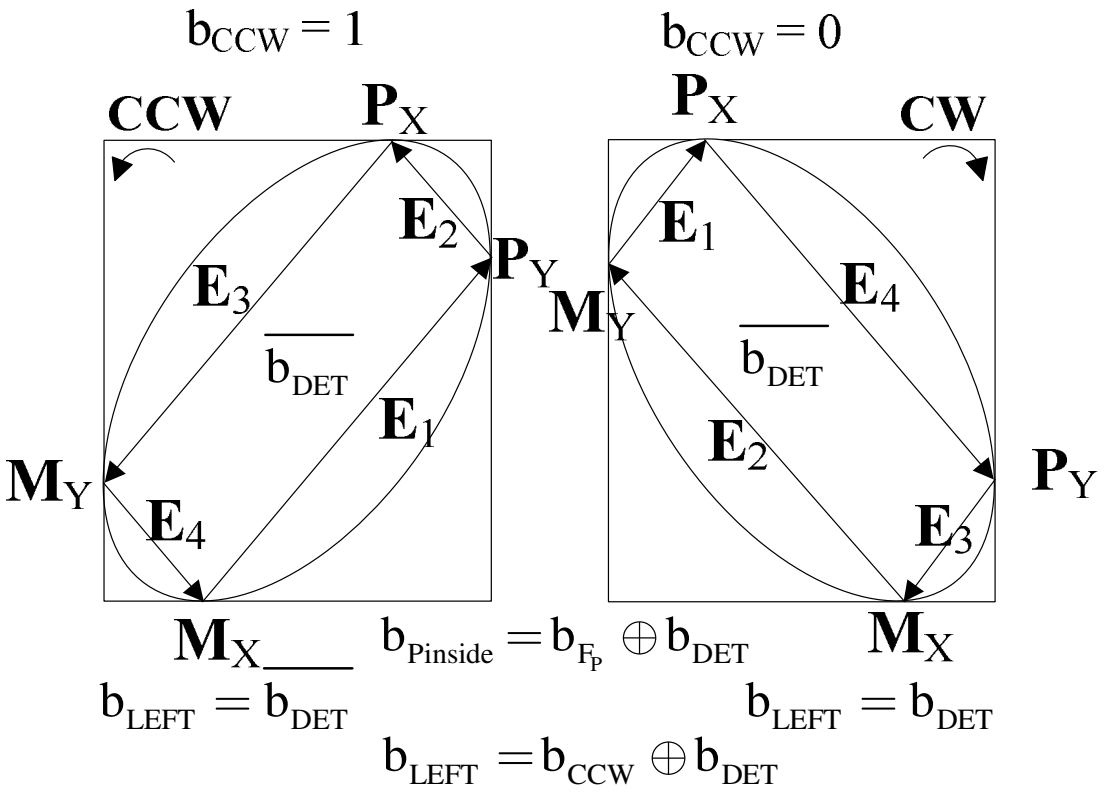

Fig. 2. The extreme vectors Ei

This paper only considers general conics defined in a point lattice, called a grid. The unit cells are squares and the minimal distance between the grid points is a rational number $\Delta$. In practical algorithms, this grid distance equals one, but using $\Delta$ clarifies and generalizes the midpoint algorithm.
We use bold characters for vectors. The three vectors $\mathbf{i}$, $\mathbf{j}$, and $\mathbf{k}$, codirectional with x, y, and z, are the mutual orthogonal unit vectors of a right-handed Cartesian coordinate system.

**Notation**

The capital letter "S" symbolizes a sign function, and the miniscule letter "b" stands for a Boolean function. FR ≜ ImplicitRegion[F[x,y]==0,{x,y}] (17). The conditional expression "result = IF( condition, value1, value2)", means that if the condition is true, then the result equals value1, else the result equals value2.

We define the Boolean-function $b_{INDEX}$ as $b_{INDEX} \triangleq \text{IF( INDEX > 0, 1, 0 )}$. So, "1" means "True" and "0" means "False", compatible with Mathematica's definitions: 1==Boole[True] and 0==Boole[False].

Most of the time, the value INDEX==0, is prefiltered out, but sometimes, to indicate what we mean, we write $b_{INDEX\geq 0} \triangleq \text{IF( INDEX} \geq 0, 1, 0 \text{ )}$.

The S-function $S_{INDEX}$ has two values $\pm 1$, because the value 0 will be filtered out; therefore we define the S-function $S_{INDEX}$ as $S_{INDEX} \triangleq \text{IF( } b_{INDEX} == 1, +1, -1 \text{ )}$.

The logical operator $\oplus$ means xor. As logical negation, we use the bar above the Boolean variable, hence $\overline{b_{INDEX}} \triangleq \neg b_{INDEX} \triangleq \text{Not}[b_{INDEX}]$.

**The Candidate Points — the Monotonic Direction — the Boolean** $b_{Lxy} = b_x \oplus b_y \oplus b_{LEFT}$

The real extreme tangent points and / or the intersection points with a bounding frame are precalculated, and rounded to the nearest grid point. These points segmentize the conic in monotonic segments, organized clockwise or counterclockwise. The start and endpoints of each i-segment, define the monotonic direction $\mathbf{E}_i \triangleq \mathbf{P}_E - \mathbf{P}_S = (x_E - x_S)\mathbf{i} + (y_E - y_S)\mathbf{j}$. The conditions $(x_E - x_S == 0)$ and $(y_E - y_S == 0)$ will be filtered out, such that the monotonic direction $(S_x, S_y)$ can be defined as,

$$b_x \triangleq \text{IF}(x_E - x_S) > 0,1,0) \quad (1), \qquad b_y \triangleq \text{IF}(y_E - y_S) > 0,1,0) \quad (2), \text{ or}$$

$$S_x \triangleq \text{IF}(x_E - x_S) > 0,1,-1) \quad (1), \qquad S_y \triangleq \text{IF}(y_E - y_S) > 0,1,-1) \quad (2).$$

Fig. 1 shows four cells with the monotonic vectors T1, T2, T3 and T4. The general monotonic vector equals $\mathbf{T}_i \triangleq S_x\Delta\mathbf{i} + S_y\Delta\mathbf{j}$ (3), with monotonic direction $(S_x, S_y)$. Point $\mathbf{P}_A$ is the actual, optimal, best selected grid point (Section 4.), and the candidate points, concurring with the monotonic vector $\mathbf{T}_i$, are the 4-connected grid points $\mathbf{P}_B$, $\mathbf{P}_C$ or the 8-connected grid points $\mathbf{P}_B$, $\mathbf{P}_C$ and $\mathbf{P}_D$. For all cases, $\mathbf{P}_B$ corresponds with a x-move and $\mathbf{P}_C$ with a y-move.

The conic $F(\mathbf{P}_A)$ with residue $F_A$ is defined as :

$$F_A \triangleq F(\mathbf{P}_A) \triangleq \mathbf{P}_A \bullet \mathbf{G}_A + W_A \triangleq [x_A \quad y_A \quad 1] \begin{bmatrix} X_A \\ Y_A \\ W_A \end{bmatrix} \triangleq [x_A \quad y_A \quad 1] \begin{bmatrix} A & D & I \\ D & B & J \\ I & J & M \end{bmatrix} \begin{bmatrix} x_A \\ y_A \\ 1 \end{bmatrix} \quad (4),$$

$$F_A \triangleq Ax_A^2 + By_A^2 + 2Dx_A y_A + 2Ix_A + 2Jy_A + M \quad (5).$$

The discriminant is $DIS \triangleq AB - D^2$. The determinant of a non-degenerated conic is $DET \triangleq \begin{vmatrix} A & D & I \\ D & B & J \\ I & J & M \end{vmatrix} \neq 0$, hence the Boolean $b_{DET}$ of the global sign of the coefficients of the non-degenerated conic is $b_{DET} = IF(DET > 0, 1, 0)$.

An exception to our notation rules is the Boolean $b_{LEFT}$ *which equals one when* $F < 0$ *on our left, when traversing the conic, else it equals zero, hence* $b_{LEFT} = IF(F(x,y) < 0, 1, 0)$ (12), but

$$S_{LEFT} = IF(b_{LEFT} == 1, +1, -1) \quad (12).$$

It is important that $b_{Lxy} \triangleq b_x \oplus b_y \oplus b_{LEFT}$ (6) and $S_{Lxy} = IF(b_{Lxy} == 1, +1, -1)$ (7) are *constants* in a monotonic segment.

The points in the candidate cell have the following properties:

$$\mathbf{P}_C \triangleq \mathbf{P}_A + S_y\Delta\mathbf{j} \quad \mathbf{P}_B \triangleq \mathbf{P}_A + S_x\Delta\mathbf{i} \quad \mathbf{P}_D \triangleq \mathbf{P}_A + \mathbf{T}_i \quad (8), \quad \mathbf{P}_C - \mathbf{P}_B = (-S_x\mathbf{i} + S_y\mathbf{j})\Delta \quad (9),$$

$$\mathbf{P}_M \triangleq \frac{\mathbf{P}_B + \mathbf{P}_C}{2} \Rightarrow \mathbf{G}_M \triangleq X_M\mathbf{i} + Y_M\mathbf{j} = \frac{\mathbf{G}_B + \mathbf{G}_C}{2} \quad (10), \quad W_M = \frac{W_B + W_C}{2} \quad (11).$$

The purpose of the algorithm is to select the optimal candidate point as fast as possible. Optimal means that the *global least square distance to the actual monotonic conic segment* is minimal.

**The Directed Polar of a Point with respect to the Conic**

For general equations, we assume that the index Z refers to an arbitrary point $\mathbf{P}_Z = x_Z\mathbf{i} + y_Z\mathbf{j}$ with gradient $\mathbf{G}_Z = X_Z\mathbf{i} + Y_Z\mathbf{j}$, then we have:

- the polar of an arbitrary point $\mathbf{P}_Z$ is the line $\mathbf{P} \bullet \mathbf{G}_Z + W_Z = 0 = (\mathbf{P} - \mathbf{P}_Z) \bullet \mathbf{G}_Z + F_Z = 0$ (13)
- the "sense of" the directed polar of $\mathbf{P}_Z$ is the vector $\mathbf{T}_{P_Z} = S_{LEFT}(\mathbf{k} \times \mathbf{G}_Z) = S_{LEFT}(-Y_Z\mathbf{i} + X_Z\mathbf{j})$ (14), with magnitude $|\mathbf{T}_{P_Z}| = |\mathbf{G}_Z|$.

**Proof**: From (12) and because the gradient points, by definition, in the direction of the greatest rate of increase of F(x,y). Cross multiplying (14) with $\times\mathbf{k}$ proves (15).

From now on we assume that the polar always points in the sense of the movement, with other words in the monotonic direction, such that the use of "sense of" can be avoided. This property of the polar will be used in section 3.

- The gradient at the pole of the directed polar $\mathbf{T}_{P_Z}$ is $\mathbf{G}_Z \triangleq X_Z\mathbf{i} + Y_Z\mathbf{j} = S_{LEFT}(\mathbf{T}_{P_Z} \times \mathbf{k})$ (15),
- The directed distance from the point $\mathbf{P}_B$ to the polar of $\mathbf{P}_Z$ is $\mathbf{r}_{BZ} = \frac{\mathbf{P}_B \bullet \mathbf{G}_Z + W_Z}{G_Z^2}\mathbf{G}_Z$ (16).

**Some important properties of conics [2, App. A.1-A.7]**

The residues $F_M$, $F_C$ and $F_B$ equal the residues of the conic in $\mathbf{P}_M$, $\mathbf{P}_C$ and $\mathbf{P}_B$.
The essential keypoints of conics are:

1. Inflexion points do not exist;
2. Every point, except the center, has a unique polar;
3. The fundamental "switching" property of the polar is $\mathbf{P}_1 \bullet \mathbf{G}_2 + W_2 == \mathbf{P}_2 \bullet \mathbf{G}_1 + W_1$ (18);
4. The conic can be divided into separate monotonic pieces. Therefore, we say that the conic can be subdivided into at most four monotonic quadrants;
5. The arithmetic mean equation is $\frac{F_2 + F_1}{2} = F\left(\frac{\mathbf{P}_2 + \mathbf{P}_1}{2}\right) + \frac{(\mathbf{P}_2 - \mathbf{P}_1)\bullet(\mathbf{G}_2 - \mathbf{G}_1)}{4}$ (19),

   the control factor $\lambda_M$ equals by definition $\lambda_M = \frac{(\mathbf{P}_C - \mathbf{P}_B)\bullet(\mathbf{G}_C - \mathbf{G}_B)}{4} = \frac{A + B - 2S_xS_yD}{4}\Delta^2$ (20),

   hence, $F_{a_M} \triangleq \frac{F_C + F_B}{2} = F_M + \lambda_M$ (21);
6. The incremental equation is $F_2 - F_1 = 2(\mathbf{P}_2 - \mathbf{P}_1)\bullet\frac{(\mathbf{G}_2 + \mathbf{G}_1)}{2}$ (22), hence,

   $F_C - F_B = 2(\mathbf{P}_C - \mathbf{P}_B)\bullet(\frac{\mathbf{G}_C + \mathbf{G}_B}{2}) = 2(\mathbf{P}_C - \mathbf{P}_B)\bullet\mathbf{G}_M$ (23);
7. The "relative or simple midpoint measurement" is $\mathbf{r}_{CM}^2 - \mathbf{r}_{BM}^2 = \frac{1}{G_M^2} F_M (F_C - F_B)$ (24),

   $\mathbf{r}_{CM}$ is the directed distance (16) from the point $\mathbf{P}_C$ to the polar of $\mathbf{P}_M$,
   $\mathbf{r}_{BM}$ is the directed distance (16) from the point $\mathbf{P}_B$ to the polar of $\mathbf{P}_M$.

## 2. INTRODUCTION

**Overview of the paper and Problem Statements**

1. **Section 4,"DTLTI-SYSTEM":** describes the optimal conditions for every incremental algorithm. The starting point must be optimal, the candidate points must belong to a monotonic segment, and the optimal candidate point can be selected "locally", if Bellman's principle of optimality holds. The dual of Bellman's principle of optimality holds for a 2D-incremental algorithm but not for 6-connected 3D incremental algorithm.
   The optimal criterion is the minimal "Global Least Square Distance" to the curve.

2. The measurement of the distance of a point $\mathbf{P}_B$ to the conic, using an advanced tool, such as Mathematica's RegionDistance$[FR, \mathbf{P}_B]$ (17) is simple, but invalid when the "tangent" (non strictly speaking the "gradient") at the footpoint of $\mathbf{P}_B$ is not conform with the monotonic direction Ti. Checking the measurement is generally complex, but even if you find an advanced checking tool, a new problem arises when there does not exists a valid measurement. The incremental methods use as measurement the midpoint method or the arithmetic mean method, also called the two-point method. Nowadays there is an agreement, that, for conics, the midpoint method is better than the two-point method. As the arithmetic mean equation (21) points out, both methods are related with a constant $\lambda_M$ and both can be invalid. In that case we say that the measurement is Out-of-Control (OoC). When we say that the midpoint method is "better", it does not mean that the two-point method is invalid, but it means that the two-point method may be inaccurate and the midpoint method may be less inaccurate, or Out-of-Accuracy (OoA).

We also show that if the grid distance is sufficient small, that inaccuracy is not as important, because the measurement can be within tolerance. So the real problem is:

2.1. How can we easily and fast detect that the measurement is invalid ?

2.1.1. **Section 3, "Algebraic OoC-condition":** the monotonic condition and the polar of the midpoint of two candidate points, can be used to check if the "polar" of the midpoint is conform with the monotonic direction Ti. We demand that the polars of the surrounding points of the midpoint of the footpoints, all have the same monotonic condition.

2.1.2. **Section 3, "Primary condition of a measurement":** the primary condition for a measurement and the gradient of the midpoint of two candidate points, predict if the measurement of the minimal distance of these points to the conic is valid. The 2$^{nd}$ condition and (see section 8.2 "Comparison with Van Aken & Novak") the 3$^{rd}$ condition for a measurement are directly consequences of the primary conditions, but the 2$^{nd}$ and 3$^{rd}$ conditions are not sufficient!

2.2. How can we continue, when all the measurements are invalid ?
**Section 6,"OoC-Rule":** The primary condition is completely new. If it is not valid, the measurement or the midpoint criterion is not used, but we show that it can be replaced with a very simple rule, which continues the algorithm and reduces the invalidity of the next measurement. The OoC-rule does not measure the distance to the conic, but it controls the digitization, such that it leaves the OoC state as soon as possible. The OoC-rule tries to correct the situation as good as possible. When we detect OoC, for all the measurements, the train is going off the rails, and the OoC-rule must put the train again on the rails or at least prevent that the train will go off the rails.

3. In stead of an advanced distance tool, we use a very simple tool, "the simple midpoint measurement". This is the core [2] of all midpoint algorithms (even for algorithm T of [1]): we measure the difference of the squared distances of two candidate points, f.e. $\mathbf{P}_C$ and $\mathbf{P}_B$ to the polar of their midpoint $\mathbf{P}_M$. Working out and simplifying this expression gives (24). When the the measurement is valid, the primary OoC-condition is true, and this expression reduces to $-S_{Lxy} F_M * 2\frac{|X_M|+|Y_M|}{G_M^2}\Delta$. Therefore, only the sign of the residue of the midpoint must be checked in a given monotonic segment.

4. **Section 5,"Relative curve measurement" theorem"** proves that the relative squared distances of two candidate points to the conic reduces to the simple midpoint measurement, provided that both measurements are valid and not Out-of-Accuracy. But inaccuracy has no effect on the digitization when the squared grid distance is smaller than half the squared worst-case tolerated tolerance range. The trick used to prove this theorem is the **"Construction of the pole $\mathbf{P}_E$" of section 2**.

5. **Section 7, "The OoA event in more details"**, describes shortly, when Out-of-Accuracy may occur. To prove the location of the OoA-segment, we must define the inner and outer of a conic, but at the same time we can simplify the formal $b_{LEFT}$-definition with a very users' friendly definition, as $b_{LEFT} = b_{CCW} \oplus b_{DET}$. In stead of $b_{LEFT}$ the new independent variable becomes $b_{CCW}$ (Fig. 2.).

## 6. Supplements:

a) Appendix 2: Average %-speed gain.
b) Appendix 3: Examples of bad and good digitalizations.
c) Appendix 4: Simple example which shows that Algorithm T of D. Knuth [1] can be OoA.
d) Berserkless 8-connected midpoint algorithm in pseudo-code (2 + 1 info pages). The modules T15 and T16 are important. The form is, intentionally, conform with algorithm T of D. Knuth [1].
e) Complete 8-connected midpoint algorithm in Mathematica-cdf format.

A user who wants supplements d or e has to send me an email.

### Relative measurements of distances, OoC and OoA

Measuring the shortest distance to a conic F(x,y)=0, ultra fast, is still a challenging problem (solution of a non-linear system) :

$\mathbf{P}_{F_C}$ and $\mathbf{P}_{F_B}$ are the footpoints of $\mathbf{P}_C$ and $\mathbf{P}_B$ on the conic, and their Euclidean distances from point $\mathbf{P}_C$ and point $\mathbf{P}_B$ to the conic are respectively $\boldsymbol{\rho}_C$ and $\boldsymbol{\rho}_B$ (Fig.3).

The midpoint of the footpoints is the point $\mathbf{P}_{F_M} = \frac{\mathbf{P}_{F_C} + \mathbf{P}_{F_B}}{2} \Rightarrow \mathbf{G}_{F_M} = \frac{\mathbf{G}_{F_C} + \mathbf{G}_{F_B}}{2}, W_{F_M} = \frac{W_{F_B} + W_{F_C}}{2}$ (25).
The footpoints must satisfy their non-linear systems,
$F_{F_C} = 0,\ \boldsymbol{\rho}_C \times \mathbf{G}_{F_C} = 0,\ \boldsymbol{\rho}_C \triangleq \mathbf{P}_C - \mathbf{P}_{F_C}$, and $F_{F_B} = 0,\ \boldsymbol{\rho}_B \times \mathbf{G}_{F_B} = 0,\ \boldsymbol{\rho}_B \triangleq \mathbf{P}_B - \mathbf{P}_{F_B}$.

With Mathematica, we can use “RegionDistance” to find the distance of a point to the conic and the minimal distance of the points $\mathbf{P}_B$, $\mathbf{P}_C$ and $\mathbf{P}_D$ to the conic. But this measurement, as the measurement with the midpoint method, can be invalid (OoC), when the candidate point does not measure the distance to *the actual monotonic conic segment.*

### Possible relative measurements

The relative distance is the difference between the squared distances from two points to some “reference”, and a relative measurement is the measurement of the relative distance. As “reference”, we will only consider a conic and the polar of a point with respect to a conic. *Limiting the relative measurement to conic curves, is not really a limitation in practice, because about every shape can be approximated by “piecewise conics”, called quadratic Bézier splines (conic splines) or squines* [1, pp. 48, pp. 181].

Replacing the “reference” with a conic or a polar gives, for two given points $\mathbf{P}_C$ and $\mathbf{P}_B$, called the candidate points, the following relative measurements:

1. The “relative curve measurement” measures the difference between the squared distances from two points to the conic, hence it measures $\boldsymbol{\rho}_C^2 - \boldsymbol{\rho}_B^2$.
2. The “relative polar measurement” measures the difference between the squared distances from two points to the polar of a special constructed pole $\mathbf{P}_E$ with respect to the conic, hence it measures $\mathbf{r}_{CE}^2 - \mathbf{r}_{BE}^2$.
3. The “relative or simple midpoint measurement” measures the difference between the squared distances from two points to the polar of the midpoint $\mathbf{P}_M$ of these points with respect to the conic, hence it measures $\mathbf{r}_{CM}^2 - \mathbf{r}_{BM}^2 \overset{(24)}{\hat{=}} \frac{1}{G_M^2} F_M (F_C - F_B)$.

All midpoint methods for conics use this criterion, hence the midpoint method measures the "relative midpoint distance".

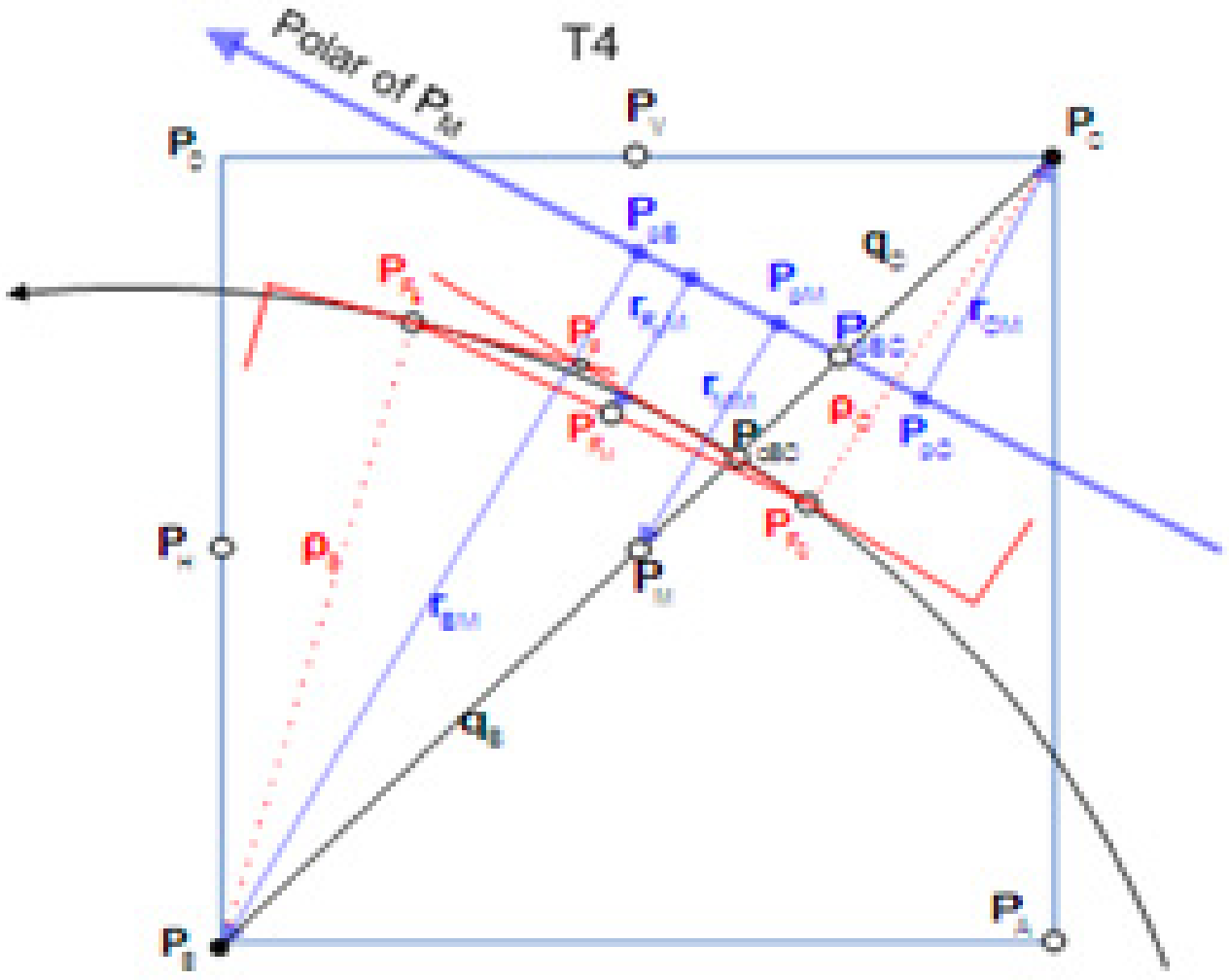

Fig. 3. Case T2

**Construction of the pole $\mathbf{P}_E$**

The tangents, in the footpoints, intersect in the pole $\mathbf{P}_F$, hence the polar $\mathbf{T}_{p_F}$ of the pole $\mathbf{P}_F$ intersects the conic in the footpoints $\mathbf{P}_{F_C}$ and ..with midpoint $\mathbf{P}_{F_M}$. The chord vector is by definition $\mathbf{L}_F \triangleq S_{L_F}\left(\mathbf{P}_{F_C}-\mathbf{P}_{F_B}\right) \in \mathbf{T}_{p_F}$ and the gradient in $\mathbf{P}_F$ is $\mathbf{G}_F$. Applying the incremental equation to the footpoints gives $2S_{L_F}\mathbf{L}_F \bullet \mathbf{G}_{F_M} = \mathbf{G}_{F_M} \bullet \mathbf{T}_{p_F} = 0 \Rightarrow \left(\mathbf{G}_{F_C} \bullet \mathbf{T}_{p_F}\right)^2 = \left(\mathbf{G}_{F_B} \bullet \mathbf{T}_{p_F}\right)^2$. In general $\left(\boldsymbol{\rho}_C \bullet \mathbf{T}_{p_F}\right)^2 \neq \left(\boldsymbol{\rho}_B \bullet \mathbf{T}_{p_F}\right)^2$, therefore we construct the pole $\mathbf{P}_E$ with gradient $\mathbf{G}_E$, such that:

1. The pivoting point of the polar $\mathbf{T}_{p_E}$ is the point $\mathbf{P}_{F_M}$.
   Hence, $\mathbf{P}_{F_M} \bullet \mathbf{G}_E + W_E = 0 \Leftrightarrow \mathbf{P}_E \bullet \mathbf{G}_{F_M} + W_{F_M} = 0$ (26).
2. We turn the polar $\mathbf{T}_{p_E}$ around the pivoting point $\mathbf{P}_{F_M}$ such that $\left(\boldsymbol{\rho}_C \bullet \mathbf{T}_{p_E}\right)^2 = \left(\boldsymbol{\rho}_B \bullet \mathbf{T}_{p_E}\right)^2$ (27).
   The tangent of the pivoting angle $\phi_E$ is a function of $\mathrm{tg}\phi_E = \dfrac{\text{function}\left(\rho_C \sin\alpha_C, \rho_B \sin\alpha_B\right)}{\text{function}\left(\rho_C \cos\alpha_C, \rho_B \cos\alpha_B\right)}$,
   with $\alpha_C = \measuredangle\left|\mathbf{P}_F-\mathbf{P}_{F_C}, \mathbf{P}_{F_B}-\mathbf{P}_{F_C}\right|$, $\alpha_B = \measuredangle\left|\mathbf{P}_F-\mathbf{P}_{F_B}, \mathbf{P}_{F_C}-\mathbf{P}_{F_B}\right|$ (28).
   It can be proved that $\left|\mathrm{tg}\phi_E\right| \leq 1$, when the measurement is valid.
3. The polar $\mathbf{T}_{p_E}$ cuts the conic in the points $\mathbf{P}_{E_C}$ and $\mathbf{P}_{E_B}$, and the tangents in these points cut in the pole $\mathbf{P}_E$ of $\mathbf{T}_{p_E}$. The chord vector is $\mathbf{L}_E \triangleq S_{L_E}\left(\mathbf{P}_{E_C}-\mathbf{P}_{E_B}\right)$ and the gradient in .. is perpendicular to the chord $\mathbf{L}_E$. The midpoints $\mathbf{P}_{E_M}$ and $\mathbf{P}_{F_M}$ of respectively the chords $\mathbf{L}_E$ and $\mathbf{L}_F$, belong to the polar of $\mathbf{P}_E$, therefore

- $\mathbf{P}_{E_M} \bullet \mathbf{G}_E + W_E = 0$ (29),
- $\mathbf{P}_{F_M} \bullet \mathbf{G}_E + W_E = 0$ (30),
- $(\mathbf{P}_{E_M} - \mathbf{P}_{F_M}) \bullet \mathbf{G}_E = 0 \Leftrightarrow \mathbf{L}_E \bullet \mathbf{G}_E = 0$.

The poles $\mathbf{P}_F$ and $\mathbf{P}_E$ belong to the polar of $\mathbf{P}_{F_M}$, and $\mathbf{P}_E - \mathbf{P}_F$ is parallel to the chord $\mathbf{L}_F = S_{L_F}(\mathbf{P}_{F_C} - \mathbf{P}_{F_B})$ or parallel to polar $\mathbf{T}_{P_F}$ of the pole $\mathbf{P}_F$.

We will use this construction in the “relative curve measurement” theorem of section 5.

**The three possible measurements ( Fig.1., Fig. 3. )**

The candidate points for every monotonic direction with optimal start point $\mathbf{P}_A$ are the points $\mathbf{P}_B$, $\mathbf{P}_C$, and $\mathbf{P}_D$ for a 8-connected digitization, or the points $\mathbf{P}_B$, $\mathbf{P}_C$ for a 4-connected digitization. As we select pairwise, we need three measurements for a 8-connected digitization and one for a 4-connected digitization. The conic is divided into separate segments, in each of which x and y are both monotonic. The Booleans of the increments of x , and y are $b_x$, $b_y$ ((1) ( 2)), and the Boolean $b_{Lxy}$ defined as $b_{Lxy} \triangleq b_x \oplus b_y \oplus b_{LEFT}$ (6), is fixed in each monotonic segment. For a monotonic conic segment, the next simple measurements are possible:

1. The M-measurement $F_M(F_C - F_B)$ using points { $\mathbf{P}_B$ , $\mathbf{P}_C$ } and their midpoint $\mathbf{P}_M \triangleq (\mathbf{P}_C + \mathbf{P}_B)/2$ with $b_{F_C\text{-}F_B} == \overline{b_{Lxy}}$ , selects point $\mathbf{P}_B$ if $b_{F_M} \oplus b_{Lxy} == 1$, else point $\mathbf{P}_C$ (31).

   The Boolean of $F_M$ is $b_{F_M}$, and the Boolean of $(F_C - F_B)$ is $\overline{b_{Lxy}}$.

   The sign of $F_M(F_C - F_B)$ is the general midpoint criterion, for a 4-connected digitization. When the “relative curve measurement” is valid and accurate, it corresponds with the Boolean expression $b_{\rho_C^2 - \rho_B^2} = b_{F_M} \oplus b_{Lxy}$ (52) meaning: if $b_{\rho_C^2 - \rho_B^2}$ is true, the midpoint method chooses point $\mathbf{P}_B$ else it chooses point $\mathbf{P}_C$.

2. The H-measurement $F_H(F_D - F_B)$ using points { $\mathbf{P}_D$ , $\mathbf{P}_B$ } and their midpoint $\mathbf{P}_H \triangleq (\mathbf{P}_D + \mathbf{P}_B)/2$ with $b_{F_D\text{-}F_B} == \overline{b_{Lxy}}$ , selects point $\mathbf{P}_B$ if $b_{F_H} \oplus b_{Lxy} == 1$, else point $\mathbf{P}_D$ (32).

3. The V-measurement $F_V(F_C - F_D)$ using points { $\mathbf{P}_C$ , $\mathbf{P}_D$ } and their midpoint $\mathbf{P}_V \triangleq (\mathbf{P}_C + \mathbf{P}_D)/2$ with $b_{F_C\text{-}F_D} == \overline{b_{Lxy}}$ , selects point $\mathbf{P}_D$ if $b_{F_V} \oplus b_{Lxy} == 1$, else point $\mathbf{P}_C$ (33).

**Proof:** The result of the “relative curve measurement” theorem of section 5.

In this paper we mostly consider the M-measurement, and we assume that the reader can apply the same reasoning to the other measurements.

**A 100 % stable hardware realization is possible**

It is our purpose to pre-design a hardware algorithm, therefore we avoid exceptions, because *simplicity favors regularity*, and therefore we will take care of all the possible “valid” measurements.

The basic forms $b_{F_M} \oplus b_{Lxy}$ , $b_{F_H} \oplus b_{Lxy}$ , $b_{F_V} \oplus b_{Lxy}$ and $\lambda_M \oplus b_{Lxy}$ (59) can be easily converted to hardware and they select the shortest distance to the conic, ultra fast. But each measurement can be invalid and an invalid measurement is not considered. The OoC-rule (59) is only applied when there are no valid measurements, therefore only the M-OoC-rule applies. *A valid measurement has always the highest priority*, except when the H-measurement selects point $\mathbf{P}_B$ and the V-measurement selects point $\mathbf{P}_C$, and the M-measurement is invalid. In that case the OoC-rule selects one of these points. In all other cases, point $\mathbf{P}_D$ gets the highest priority. When all the measurements are invalid, we apply the OoC-Rule, which selects the most stable point out of { $\mathbf{P}_B$ , $\mathbf{P}_C$ }.

## 3. Primary OoC-condition of a Measurement

**Algebraic OoC-condition**

The midpoint measurement is OoC, *if the sense of the direction of the digitization is not conform to the sense of the monotonic direction.*

The monotonic vector $\mathbf{T}_i$ measures the monotonic direction, and the sense of the directed polar $\mathbf{T}_{P_Z}$ of the midpoint $\mathbf{P}_Z$, measures the sense of the direction of the digitization. Hence, the polar $\mathbf{T}_{P_Z}$ of $\mathbf{P}_Z$ must be *monotonically equal to the monotonic vector* $\mathbf{T}_i$, notated as $\mathbf{T}_{P_Z} \stackrel{\leftarrow}{\rightharpoonup} \mathbf{T}_i$ (or simply said conform with the monotonic vector $\mathbf{T}_i$ ). With $\mathbf{A \bullet B} = A_x * B_x + A_y * B_y$ , and $\langle \mathbf{A \bullet B} \rangle > 0$ defined as $A_x * B_x > 0$ and $A_y * B_y > 0$ , the algebraic OoC-condition for a valid measurement, associated with the midpoint $\mathbf{P}_Z$ is

$$\mathbf{T}_{P_Z} \stackrel{\leftarrow}{\rightharpoonup} \mathbf{T}_i \Leftrightarrow \langle \mathbf{T}_{P_Z} \bullet \mathbf{T}_i \rangle > 0 \quad (34).$$

The notation between quotes,
"$\mathbf{T}_{P_B}$ , $\mathbf{T}_{P_C}$ , $\mathbf{T}_{P_D}$ , $\mathbf{T}_{P_M}$ , $\mathbf{T}_{P_V}$ , $\mathbf{T}_{P_H}$ , the polars of the surrounding points of the midpoint of the footpoints,... $\stackrel{\leftarrow}{\rightharpoonup} \mathbf{T}_i$ " indicates, in detail, which polars are involved , but it always means that the measurement(s) corresponding with the poles of the polars, must be valid.

**Primary OoC-conditions in Boolean form**

With (1), (2) and the Boolean .. defined as $b_{Lxy} \triangleq b_x \oplus b_y \oplus b_{LEFT}$ , <u>the necessary and sufficient conditions</u> that a midpoint measurement is valid are:

a. for the measurement using the midpoint $\mathbf{P}_M$ : $b_M = b_{M_1} \wedge b_{M_2} == 1$,
with $b_{M_1} = \overline{b_y \oplus b_{Y_M} \oplus b_{Lxy}}$ , $b_{M_2} = b_x \oplus b_{X_M} \oplus b_{Lxy}$ (35);

b. for the measurement using the midpoint $\mathbf{P}_H$ : $b_H = b_{H_1} \wedge b_{H_2} == 1$,
with $b_{H_1} = \overline{b_y \oplus b_{Y_H} \oplus b_{Lxy}}$ , $b_{H_2} = b_x \oplus b_{X_H} \oplus b_{Lxy}$ (36);

c. for the measurement using the midpoint $\mathbf{P}_V$ : $b_V = b_{V_1} \wedge b_{V_2} == 1$,
with $b_{V_1} = \overline{b_y \oplus b_{Y_V} \oplus b_{Lxy}}$ , $b_{V_2} = b_x \oplus b_{X_V} \oplus b_{Lxy}$ (37).

We call these conditions the *primary OoC-conditions*. These conditions can be checked for each measurement, and they must be valid for all other known points, but also for all unknown poles, such as $\mathbf{P}_E$, $\mathbf{P}_F$, $\mathbf{P}_Z$, etcetera.

**Proof:** We will only prove the conditions for the pole $\mathbf{P}_E$, hence we have to prove $b_E = b_{E_1} = b_{E_2} = 1$, with $b_{E_1} = \overline{b_y \oplus b_{Y_E} \oplus b_{Lxy}}$, $b_{E_2} = b_x \oplus b_{X_E} \oplus b_{Lxy}$, and $b_E = b_{E_1} \,\&\, b_{E_2}$; or in S-form:

$S_{E_1} = -S_y * S_{Y_E} * S_{Lxy} = 1$, $S_{E_2} = S_x * S_{X_E} * S_{Lxy} = 1$, hence $S_E = S_{E_1} * S_{E_2} = 1$.

$$\mathbf{T}_{P_E} \bullet \mathbf{T}_i = S_{LEFT}(\mathbf{k} \times \mathbf{G}_E) \bullet (S_x \mathbf{i} + S_y \mathbf{j})\Delta = S_{Lxy} S_x S_y \left[ (\mathbf{i} \times \mathbf{k}) \bullet \mathbf{G}_E * S_x + (\mathbf{j} \times \mathbf{k}) \bullet \mathbf{G}_E * S_y \right] \Delta$$
$$= \left[ S_{Lxy} S_y (-\mathbf{j}) \bullet \mathbf{G}_E + S_{Lxy} S_x (\mathbf{i}) \bullet \mathbf{G}_E \right] \Delta = -S_{Lxy} S_y S_{Y_E} |Y_E| \Delta + S_{Lxy} S_x S_{X_E} |X_E| \Delta$$
$$= S_{E_1} |Y_E| \Delta + S_{E_2} |X_E| \Delta .$$

The polar of $\mathbf{P}_E$ is monotonically equal to the monotonic vector $\mathbf{T}_i$, when $\langle \mathbf{T}_{P_Z} \bullet \mathbf{T}_i \rangle > 0$, therefore $S_{E_1}$, $S_{E_2}$ and $S_E$ must equal one.

#### Necessary but Insufficient secondary conditions in Boolean form

When the measurements are valid then

a. for the measurement using the midpoint $\mathbf{P}_M$: $b_{(\mathbf{P}_C - \mathbf{P}_B) \bullet \mathbf{G}_E} = b_{(\mathbf{P}_C - \mathbf{P}_B) \bullet \mathbf{G}_M} = b_{F_C - F_B} == \overline{b_{Lxy}}$ (38).

b. for the measurement using the midpoint $\mathbf{P}_H$: $b_{(\mathbf{P}_D - \mathbf{P}_B) \bullet \mathbf{G}_E} = b_{(\mathbf{P}_D - \mathbf{P}_B) \bullet \mathbf{G}_H} = b_{F_D - F_B} == \overline{b_{Lxy}}$ (39).

c. for the measurement using the midpoint $\mathbf{P}_V$: $b_{(\mathbf{P}_C - \mathbf{P}_D) \bullet \mathbf{G}_E} = b_{(\mathbf{P}_C - \mathbf{P}_D) \bullet \mathbf{G}_V} = b_{F_C - F_D} == \overline{b_{Lxy}}$ (40).

We call these conditions the *secondary OoC-conditions*. Each measurement will apply this condition.

The secondary OoC-conditions are necessary but not sufficient.

*We can replace $\mathbf{G}_E$ with $\mathbf{G}_M$, $\mathbf{G}_H$, $\mathbf{G}_V$, $\mathbf{G}_C$, $\mathbf{G}_B$, $\mathbf{G}_Z$, etc.*

**Proof:** We will only prove (38) with index E:
$(\mathbf{P}_C - \mathbf{P}_B) \bullet \mathbf{G}_E = (S_y \mathbf{j} - S_x \mathbf{i}) \bullet \mathbf{G}_E \Delta = (S_y S_{Y_E} |Y_E| - S_x S_{X_E} |X_E|)\Delta$. Applying the primary OoC-conditions $S_y * S_{Y_E} = -S_{Lxy}$ and $S_x * S_{X_E} = S_{Lxy}$ gives, $(\mathbf{P}_C - \mathbf{P}_B) \bullet \mathbf{G}_E = -S_{Lxy}\left(|Y_E| + |X_E|\right)\Delta$. Proving with index M gives $(\mathbf{P}_C - \mathbf{P}_B) \bullet \mathbf{G}_M = -S_{Lxy}\left(|Y_M| + |X_M|\right)\Delta$. Applying the incremental equation (23) proves (38).

## 4. DTLTI-SYSTEM

Looking at (8, 9), the digitizing of 2D-curves can be seen as a Deterministic Discrete-Time Linear Time-Invariant System [10]. The digitized point at time $t_n$ corresponds to the stage n. For a digitized 2D-curve, the state difference equation is $\begin{bmatrix} x_{n+1} \\ y_{n+1} \end{bmatrix} = \begin{bmatrix} x_n \\ y_n \end{bmatrix} + \begin{bmatrix} S_x & 0 \\ 0 & S_y \end{bmatrix} \begin{bmatrix} u_{nx} \\ u_{ny} \end{bmatrix}$, and the inputs are $u_{nx}, u_{ny} \in \{0, \Delta\}$ and $u_{nx} + u_{ny} = \Delta$ for a 4-connected 2D-curve, and $u_{nx} + u_{ny} = \Delta$ or $2\Delta$ for a 8-connected 2D-curve. This system is time-invariant when $\begin{bmatrix} S_x & 0 \\ 0 & S_y \end{bmatrix}$ is independent of the time, and this is the case, as long as the stages belong to their monotonic quadrant. Therefore, *the monotonic part of a digitized curve can be modeled as a DTLTI-system*, the input can be considered as the set of feasible decisions $U_n = \{\mathbf{u}_n\} = \{S_x\Delta\mathbf{i}, S_y\Delta\mathbf{j}, \Delta(S_x\mathbf{i} + S_y\mathbf{j})\}$ and the state equation (or system function) becomes, $\mathbf{P}_{n+1} = \mathbf{P}_n + \mathbf{u}_n$. The states equal the grid position vectors of the discrete 2D-curve, $\mathbf{P}_n$ equals $\mathbf{P}_A$, and $\mathbf{P}_{n+1}$ equals $\mathbf{P}_B$, $\mathbf{P}_C$ or $\mathbf{P}_D$. The partial trajectory $\mathbf{T}_0^r$ equals $(\mathbf{P}_0, \mathbf{P}_1, \mathbf{P}_2, \cdots, \mathbf{P}_r) = (\mathbf{P}_0, \mathbf{u}_0, \mathbf{u}_1, \cdots, \mathbf{u}_{r-1}) = (\mathbf{P}_0, \mathscr{P}_0^{r-1}) = \mathbf{T}_0^r$, so it depends only on the initial state $\mathbf{P}_0$ and the policy $\mathscr{P}_0^{r-1}$ of the first r decisions $(\mathbf{u}_0, \mathbf{u}_1, \cdots, \mathbf{u}_{r-1})$. The cost per stage $c_n(\mathbf{P}_n, \mathbf{u}_n)$ is the criterion used to select $\mathbf{P}_n$ from the candidate points $\mathbf{P}_B$, $\mathbf{P}_C$ and $\mathbf{P}_D$ given the point $\mathbf{P}_A = \mathbf{P}_{n-1}$, and the possible moves $U_{n-1} = \{S_x\Delta\mathbf{i}, S_x\Delta\mathbf{j}, \Delta(S_x\mathbf{i} + S_y\mathbf{j})\}$, hence the cost per stage $c_n(\mathbf{P}_n, \mathbf{u}_n)$ is independent of the decision $\mathbf{u}_n$ and depends only on the forward partial trajectory $\mathbf{T}_0^n = (\mathbf{P}_0, \mathbf{P}_1, \mathbf{P}_2, \cdots, \mathbf{P}_n) = (\mathbf{P}_0, \mathbf{u}_0, \mathbf{u}_1, \cdots, \mathbf{u}_{n-1})$. We assume, that the Least Square Distance is the criterion $c_n(\mathbf{P}_n, \mathbf{u}_n)$ used for digitizing the conic. The set of feasible decisions is independent of the stage, provided that the stages remain in their monotonic quadrant. Therefore, *the system is time-invariant and deterministic as long as the stages belong to their monotonic quadrant*. The objective is to find a complete trajectory $\mathbf{T}$ that minimizes the cost of the complete trajectory

$$V(\mathbf{T}) = V_0^N(\mathbf{T}_0^N) = \sum_{n=0}^{n=N} c_n(\mathbf{P}_n, \mathbf{u}_n).$$

*For a time-invariant deterministic dynamic system*, the recursive procedure can be based on a *forward* induction process, where the first stage to be solved is the initial stage of the problem, and problems are solved moving forward one stage at a time, until all stages are included [11],[12]. ." Hence, *the starting point* $\mathbf{P}_A$ *of (8) must be optimal* ! The basis of the forward recursive optimization procedure is a dual to Bellman's statement: "An optimal policy has the property that, whatever the ensuing state and decisions are, the preceding decisions must constitute an optimal policy with respect to the state existing before the last decision.
*If the dual of Bellman's' principle of optimality holds, then*

$$J_0^{r+1} = \min[V_0^r(\mathbf{T}_0^r)] + \min[c_{r+1}(\mathbf{P}_{r+1}, \mathbf{u}_{r+1})]] == J_0^r + \underbrace{\min[c_{r+1}(\mathbf{P}_{r+1}, \mathbf{u}_{r+1})]]}_{\text{local}}$$

$\Leftrightarrow$ **The global minimum equals the sum of the local minima**.

The solution of the DP-problem does not say how we have to select the best point or how we have to find the best decision vector. It just says, *if Bellman's principle or its dual holds, then you have a local problem, and if you can find a solution for that local problem, that solution is also valid for the global problem.* It is clear that the principles hold for a 4- or 8-connected 2D-curve, and a 26-connected 3D-curve, but not for a 6-connected 3D-curve! Therefore, the Tripod 6-Connected 3D Line algorithm [13] is not global optimal.

## 5. "RELATIVE CURVE MEASUREMENT" THEOREM

The "relative curve measurement" theorem proves that

$$\boldsymbol{\rho}_C^2 - \boldsymbol{\rho}_B^2 = \frac{2}{G_E^2}(1-\varepsilon_E)\tau_E F_M(\mathbf{P}_C - \mathbf{P}_B)\bullet\mathbf{G}_E \quad \text{or} \quad b_{\rho_C^2-\rho_B^2} = b_{F_M} \oplus b_{(\mathbf{P}_C-\mathbf{P}_B)\bullet\mathbf{G}_E} \oplus b_{\tau_E} \oplus b_{(1-\varepsilon_E)}.$$

When a measurement is valid, the parameter $\varepsilon_E$ is always smaller than one, and when the measurement is accurate, we have $\tau_E > 0$, hence $b_{|\rho_C|>|\rho_B|} = b_{F_M} \oplus b_{Lxy}$.

**Proof**: In section 2, we constructed the pole $\mathbf{P}_E$ such that $\left(\boldsymbol{\rho}_C\bullet\mathbf{T}_{p_E}\right)^2=\left(\boldsymbol{\rho}_B\bullet\mathbf{T}_{p_E}\right)^2$.

Using the identity $\left(\boldsymbol{\rho}_i\bullet\mathbf{T}_{pE}\right)^2=\boldsymbol{\rho}_i^2*\mathbf{G}_E^2-\left(\boldsymbol{\rho}_i\bullet\mathbf{G}_E\right)^2$, from appendix 1, it is now easy to calculate $\boldsymbol{\rho}_C^2-\boldsymbol{\rho}_B^2$ because $\boldsymbol{\rho}_C^2*\mathbf{G}_E^2-\left(\boldsymbol{\rho}_C\bullet\mathbf{G}_E\right)^2=\boldsymbol{\rho}_B^2*\mathbf{G}_E^2-\left(\boldsymbol{\rho}_B\bullet\mathbf{G}_E\right)^2$.

Therefore $\left(\boldsymbol{\rho}_C^2-\boldsymbol{\rho}_B^2\right)*\mathbf{G}_E^2$ equals

$$\left[\underbrace{(\boldsymbol{\rho}_C+\boldsymbol{\rho}_B)}_{\mathbf{P}_C+\mathbf{P}_B-(\mathbf{P}_{F_C}+\mathbf{P}_{F_B})}\bullet\mathbf{G}_E\right]*\left[\underbrace{(\boldsymbol{\rho}_C-\boldsymbol{\rho}_B)}_{\mathbf{P}_C-\mathbf{P}_B-(\mathbf{P}_{F_C}-\mathbf{P}_{F_B})}\bullet\mathbf{G}_E\right]=2\left[\mathbf{P}_M\bullet\mathbf{G}_E-\mathbf{P}_{F_M}\bullet\mathbf{G}_E\right]*\left[(\mathbf{P}_C-\mathbf{P}_B)\bullet\mathbf{G}_E-(\mathbf{P}_{F_C}-\mathbf{P}_{F_B})\bullet\mathbf{G}_E\right]$$

$$=2\left[\underbrace{\mathbf{P}_M\bullet\mathbf{G}_E+W_E}_{\underbrace{\mathbf{P}_E\bullet\mathbf{G}_M+W_M}_{\tau_E F_M}}-\underbrace{\left(\mathbf{P}_{F_M}\bullet\mathbf{G}_E+W_E\right)}_{0}\right]*\left[\left[(\mathbf{P}_C-\mathbf{P}_B)\bullet\mathbf{G}_E-\underbrace{\left(\left(\mathbf{P}_{F_C}-\mathbf{P}_{F_B}\right)\bullet\mathbf{G}_E\right)}_{\varepsilon_E(\mathbf{P}_C-\mathbf{P}_B)\bullet\mathbf{G}_E}\right]\right].$$

Hence, the "relative curve measurement" is $\boldsymbol{\rho}_C^2-\boldsymbol{\rho}_B^2=\frac{2}{G_E^2}(1-\varepsilon_E)\tau_E F_M(\mathbf{P}_C-\mathbf{P}_B)\bullet\mathbf{G}_E$ (41).

These parameters have only sense when the measurement is valid, and they equal

$\varepsilon_E\triangleq\frac{(\mathbf{P}_{FC}-\mathbf{P}_{FB})\bullet\mathbf{G}_E}{(\mathbf{P}_C-\mathbf{P}_B)\bullet\mathbf{G}_E}\triangleq S_{L_F}\frac{\mathbf{L}_F\bullet\mathbf{G}_E}{(\mathbf{P}_C-\mathbf{P}_B)\bullet\mathbf{G}_E}$ (42), and $\tau_E\triangleq\frac{\mathbf{P}_E\bullet\mathbf{G}_M+W_M}{F_M}$ (43).

The directed distance of the point $\mathbf{P}_{F_C}$ to the polar of $\mathbf{P}_E$ is $\mathbf{r}_{F_CE}=\frac{\mathbf{P}_{F_C}\bullet\mathbf{G}_E+W_E}{G_E^2}\mathbf{G}_E$. Therefore,

- $\mathbf{r}_{F_CE}+\mathbf{r}_{F_BE}=\frac{(\mathbf{P}_{F_C}+\mathbf{P}_{F_B})\bullet\mathbf{G}_E+2W_E}{G_E^2}\mathbf{G}_E=2\frac{\mathbf{P}_{F_M}\bullet\mathbf{G}_E+W_E}{G_E^2}\mathbf{G}_E=0$ (44),

  because the pivoting point $\mathbf{P}_{F_M}$ belongs to the polar of $\mathbf{P}_E$ and $\mathbf{P}_F$.
- $\mathbf{r}_{F_CE}-\mathbf{r}_{F_BE}=\varepsilon_E(\mathbf{r}_{CE}-\mathbf{r}_{BE})$ (45) and you may expect that $\varepsilon_E$ will be very small, as $\lim_{E\to F}\varepsilon_E$ equals 0, because $\mathbf{L}_F\bullet\mathbf{G}_F==0$. Algebraic, it is rather complex to prove, that $b_{(1-\varepsilon_E)}==1$. One has to prove that $L_F<\sqrt{2}*\Delta$ and that the $|\text{pivoting angle}|=$ $|\phi_E|=\left|\measuredangle(\mathbf{T}_{p_E},\mathbf{T}_{p_F})\leq\right|=\left|\measuredangle(\mathbf{G}_F,\mathbf{G}_E)\right|\leq\pi/4$ . Therefore, we decided to cancel that proof in this paper.
- $\mathbf{r}_{CE}+\mathbf{r}_{BE}=\frac{(\mathbf{P}_C+\mathbf{P}_B)\bullet\mathbf{G}_E+2W_E}{G_E^2}\mathbf{G}_E=2\frac{\mathbf{P}_M\bullet\mathbf{G}_E+W_E}{G_E^2}\mathbf{G}_E=2\frac{\mathbf{P}_E\bullet\mathbf{G}_M+W_M}{G_E^2}\mathbf{G}_E$ (46).

  Equation (46) is the result of applying the "switching" property for polars (18). We know that the residue in the midpoint $\mathbf{P}_M$ equals $F_M=\mathbf{P}_M\bullet\mathbf{G}_M+W_M$, hence the natural choice is to link $\mathbf{P}_E\bullet\mathbf{G}_M+W_M$ to $\mathbf{P}_M\bullet\mathbf{G}_M+W_M$ , using the parameter $\tau_E\triangleq\frac{\mathbf{P}_E\bullet\mathbf{G}_M+W_M}{F_M}$.

  The parameter $\tau_E$ becomes one for $\tau_M$.

Therefore (41) defines different measurements in function of the parameters $\varepsilon_E$ and $\tau_E$:

- the "relative midpoint measurement" measures the difference between the squared distances from the candidate points $\mathbf{P}_C$ and $\mathbf{P}_B$ to the polar of $\mathbf{P}_M$, and it equals

  $\mathbf{r}_{CM}^2-\mathbf{r}_{BM}^2=\frac{2}{G_M^2}\tau_M F_M(\mathbf{P}_C-\mathbf{P}_B)\bullet\mathbf{G}_M=\frac{1}{G_M^2}F_M(F_C-F_B)$ (47),

  This measurement is accurate, and has no tolerance, but it can be invalid (OoC);

- the "relative polar measurement" $\mathbf{r}_{CE}^2 - \mathbf{r}_{BE}^2 = \frac{2}{G_E^2}\tau_E F_M(\mathbf{P}_C - \mathbf{P}_B)\bullet\mathbf{G}_E$ (48)

  measures the difference between the squared distances from the candidate points $\mathbf{P}_C$ and $\mathbf{P}_B$ to the polar of $\mathbf{P}_E$ with respect to the conic;
- the "relative curve measurement" $\boldsymbol{\rho}_C^2 - \boldsymbol{\rho}_B^2 = \frac{2}{G_E^2}(1-\varepsilon_E)\tau_E F_M(\mathbf{P}_C - \mathbf{P}_B)\bullet\mathbf{G}_E$ measures the difference between the squared distances from the candidate points $\mathbf{P}_C$ and $\mathbf{P}_B$ to the conic.

When the measurement is valid, we have $b_{(1-\varepsilon_E)} == 1$ and the secondary OoC-conditions $b_{(\mathbf{P}_C-\mathbf{P}_B)\bullet\mathbf{G}_E} = \overline{b_{Lxy}}$ , etc. hold. The parameters $\tau_E$ and $\tau_F$ are responsible for Out-of-Accuracy. From the definition of OoA, we have $b_{\tau_E} = 1$ when the measurement is accurate, and $b_{\tau_E} = 0$ when the measurement is Out-of-Accuracy (OoA).

When the measurement is valid, we have,

- $b_{(\mathbf{P}_C-\mathbf{P}_B)\bullet\mathbf{G}_E} = \overline{b_{Lxy}}$ , etc. (49),
- $b_{(1-\varepsilon_E)} = 1$ (50),
- $b_{\rho_C^2-\rho_B^2} = b_{F_M} \oplus b_{Lxy} \oplus \overline{b_{\tau_E}}$ (51).

If the measurement is valid and accurate then $b_{\tau_E} = 1$. Therefore, when the measurement is valid and accurate, we have $b_{|\rho_C|>|\rho_B|} = b_{F_M} \oplus b_{Lxy}$ (52).

So, the "relative curve measurement" reduces to the "relative simple midpoint measurement" provided that the measurement is valid and accurate.

**Tolerance of the measurement**

When the measurement is inaccurate it can be within the tolerance. The worst-case tolerance range is defined as the maximal distance between the candidate points $\left(\Delta\sqrt{2}\right)$, and the tolerance equals half the tolerance range.
If the measurement is inaccurate ( $b_{\tau_E} = 0$ ) then we have $b_{\rho_C^2-\rho_B^2} = \overline{b_{F_M} \oplus b_{Lxy}}$ , the midpoint method is wrong, and the measurement is "Out-of-Accuracy" (OoA). But the measurement is within tolerance if

$$\Delta \le \frac{\text{worst-case tolerance range which can be tolerated}}{\sqrt{2}} = \sqrt{2} * \text{worst-case tolerance} \qquad (53).$$

**Avoiding Aliasing using the (Nyquist-Shannon) Sampling Theorem:**

The radius of curvature of a conic equals $R_{Cur} = \frac{G^3}{DET}$, and the sampling theorem states that $\left|\frac{R_{Cur}}{\Delta}\right| \ge 1$ (common minimum values are 5 à 10, say $n_S$ ). Therefore the grid-distance must be smaller then Minimum( $\sqrt{2} * \text{worst-case tolerance}$ , Minimum( $\left|\frac{R_{Cur}}{n_S}\right|$ ) ).

A circle with a radius equal to the grid-distance digitizes as a square!

**Two-point method or midpoint method**

We show in section 7, that OoA happens once in a while, when the midpoint is, in practice, inside and very, very near to the conic. In that case we have,

- $|F_M| < |F_{E_M}| = |\lambda_{E_M}|$ (54),
- $\| \boldsymbol{\rho}_C| - |\boldsymbol{\rho}_B\| \ll$ Tolerance,
- $|\boldsymbol{\rho}_C| \le$ tolerance and $|\boldsymbol{\rho}_B| \le$ tolerance ,
- $\tau_E F_M = \tau_{a_M} F_{a_M} \overset{(20)}{=} \tau_{a_M}(F_M + \lambda_M)$ .

Hence, $b_{\tau_E F_M} = b_{\tau_{a_M} F_{a_M}}$ and *that resolves the long existing conflict between the midpoint method and the two-point method.* In practice, if one detects that the sign of $F_M$ and $F_{a_M}$ are different then one may have OoA. Section 7 proves that this way of detecting OoA, has many false signals. Simply said the midpoint method is better than the two-point method (for conics).

## 6. OoC-Rule

Beforehand, one calculates $\Lambda_{\lambda_M} = A + B - 2S_x S_y D$ (55), $b_{\lambda_M} = \text{IF}(\Lambda_{\lambda_M} \ge 0, 1, 0)$ (56), and $b_{\lambda_M} = \text{IF}(\Lambda_{\lambda_M} = 0, \overline{b_{F_M}}, b_{\lambda_M})$ (56).

The update equations from $\mathbf{P}_M$ to $\hat{\mathbf{P}}_M$ are: $\hat{X}_M = X_M + (b_{xmove} S_x A + b_{ymove} S_y D)\Delta$ (57), and $\hat{Y}_M = Y_M + (b_{xmove} S_x D + b_{ymove} S_y B)\Delta$ (58).

The OoC-Rule $b_{xmove} = \overline{b_\lambda \oplus b_{Lxy}}$ , $b_{ymove} = b_\lambda \oplus b_{Lxy}$ must be applied when $b_M = b_H = b_V = 0$, viz. when all the measurements are invalid. For the proof, we focus on the M-measurement.

**Proof:**

From the incremental equation (23) and (9), we have $F_C - F_B = 2(S_y Y_M - S_x X_M)\Delta$ .

When the primary condition is true then $F_C - F_B = -2S_{Lxy}(|Y_M| + |X_M|)\Delta$, and if $S_{Lxy}$ equals +1 then $S_y Y_M - S_x X_M < 0$ . But when $S_y Y_M - S_x X_M \ge 0$ and $S_{Lxy}$ equals +1, then the primary condition is not true and the best solution is to make $S_y Y_M - S_x X_M$ as negative as possible without applying the invalid M-measurement. Otherwise, when the primary condition is true and if $S_{Lxy}$ equals -1 then $S_y Y_M - S_x X_M > 0$. But when $S_y Y_M - S_x X_M \le 0$ and $S_{Lxy}$ equals -1, then the primary condition is not true and the best solution is to make $S_y Y_M - S_x X_M$ as positive as possible without applying the invalid midpoint method. Therefore, we have the next cases:

▹ **Case 1**, $b_{Lxy} = 0$ and $b_M = 0$:

From position $\mathbf{P}_A$ , we have to find a single step, which brings us to the new position $\hat{\mathbf{P}}_A$ , and the new difference becomes $(S_y \hat{Y}_M - S_x \hat{X}_M)$. The only thing we can do, is to choose the step, which makes $S_y \hat{Y}_M - S_x \hat{X}_M$ the most positive. Therefore, $S_y Y_M - S_x X_M < S_y \hat{Y}_M - S_x \hat{X}_M$ or $0 < S_y(\hat{Y}_M - Y_M) - S_x(\hat{X}_M - X_M)$ . Using the update equations gives,

- for a x-step $0 < S_x S_y D - A$ , because $\hat{Y}_M - Y_M = S_x D\Delta$ and $\hat{X}_M - X_M = S_x A\Delta$ ;
- for a y-step $0 < B - S_x S_y D$ , because $\hat{Y}_M - Y_M = S_y B\Delta$ and $\hat{X}_M - X_M = S_y D\Delta$ .

Therefore, given $b_{Lxy} = 0$, a x-step is better than a y-step if $B - S_x S_y D < S_x S_y D - A$ or if $\underbrace{A + B - 2S_x S_y D}_{\frac{4\lambda_M}{\Delta^2}} < 0$, or $b_{xmove} = \overline{b_{\lambda_M} \oplus b_{Lxy}}$ and $b_{ymove} = b_{\lambda_M} \oplus b_{Lxy}$ (59).

▷ **Case 2**, $b_{Lxy} = 1$ and $b_M = 0$:

From position $\mathbf{P}_A$, we have to find a single step, which brings us to the new position $\hat{\mathbf{P}}_A$, and the new difference becomes $(S_y \hat{Y}_M - S_x \hat{X}_M)$. The only thing we can do, is to choose the step, which makes $S_y \hat{Y}_M - S_x \hat{X}_M$ the most negative. Therefore, $S_y \hat{Y}_M - S_x \hat{X}_M < S_y Y_M - S_x X_M$ or $S_y (\hat{Y}_M - Y_M) - S_x (\hat{X}_M - X_M) < 0$. Using the update equations gives,

- for a x-step $S_x S_y D - A < 0$, because $\hat{Y}_M - Y_M = S_x D\Delta$ and $\hat{X}_M - X_M = S_x A\Delta$;
- for a y-step $B - S_x S_y D < 0$, because $\hat{Y}_M - Y_M = S_y B\Delta$ and $\hat{X}_M - X_M = S_y D\Delta$.

Therefore, given $b_{Lxy} = 1$, a x-step is better than a y-step if $S_x S_y D - A < B - S_x S_y D$ or if $\underbrace{A + B - 2S_x S_y D}_{\frac{4\lambda_M}{\Delta^2}} > 0$, or $b_{xmove} = \overline{b_{\lambda_M} \oplus b_{Lxy}}$ and $b_{ymove} = b_{\lambda_M} \oplus b_{Lxy}$ (59).

The control factor $\lambda_M$ can become zero for a parabola and a hyperbola. In that case, it does not matter if we make a x- or a y-move, and we can just as well make a move that corresponds with the measurement. Hence, $b_{xmove} = b_{F_M} \oplus b_{Lxy}$ or $b_{\lambda_M} = \mathrm{IF}(\Lambda_{\lambda_M} = 0, \overline{b_{F_M}}, b_{\lambda_M})$.

## 7. THE OoA EVENT IN MORE DETAILS

To explain the OoA event, and to find the location of the midpoints $\{\mathbf{P}_M, \mathbf{P}_H, \mathbf{P}_V\}$, we need to know the sign of the inside of a conic and the sign of the residue at a point to the left or the right of a polar.

**Inside and outside of a conic**

Determining the inside of a curve can be very complex. Knuth [1, pp. 44] used the Jordan curve theorem, we use the topological definition: a point is inside a conic if the conic is concave, seen from that point. Therefore the center of an ellipse is inside, and the center of a hyperbola (intersection of the asymptotes) is at the outside of the conic. From [9], the residue $F_{CT}$ at the center of a conic equals $\frac{DET}{DIS}$, therefore the Boolean of the residue of the center of an ellipse or hyperbola is $b_{F_{CT}} = \overline{b_{DIS} \oplus b_{DET}}$ (60).

The center of a parabola is at infinity, but empirical it is, as the center of a hyperbola, at the outside. So, we can say that the center of a every conic is inside, if $DIS > 0$ and outside, if $DIS \le 0$.

Therefore the algebraic definition of inside is $b_{Pinside} = \overline{b_{F_P} \oplus b_{DET}}$ (61) because

- $b_{DIS} = \overline{b_{F_{CT}} \oplus b_{DET}}$,
- $b_{P_{CT}inside} == b_{DIS}$, and
- $b_{F_P} == b_{F_{CT}}$ as long as the point and the center are inside or outside the conic.

*The sign of the residue of a point, determines if the point is inside or outside the conic, but it also depends on the sign of the determinant.* We need the inside-definition to locate the OoA-segment, but at the same time, it gives a user's friendly approach of defining $b_{LEFT}$ : traveling clockwise or counterclockwise is the person's view from the inside. Using this approach, it is easy to prove that $b_{LEFT}$ equals $b_{LEFT} = b_{CCW} \oplus b_{DET}$ (62).

**The polar divides the plane in a + and - half plane**

If $\mathbf{P}$ and $\mathbf{P}_E$ are at the same side of the polar line of $\mathbf{P}_E$ then $b_{(\mathbf{P}-\mathbf{P}_{E_M})\bullet\mathbf{G}_E} = b_{F_E}$ (63) else $b_{(\mathbf{P}-\mathbf{P}_{E_M})\bullet\mathbf{G}_E} = \overline{b_{F_E}}$ and $b_{(\mathbf{P}-\mathbf{P}_{F_M})\bullet\mathbf{G}_E} = b_{F_E}$ (64) else $b_{(\mathbf{P}-\mathbf{P}_{F_M})\bullet\mathbf{G}_E} = \overline{b_{F_E}}$ with $b_{F_E}$ the Boolean of the residue $F_E = F(\mathbf{P}_E)$ .

**The location of the midpoint $\mathbf{P}_M$ defines the sign of $\tau_E$**

From (43), the switching property (18), and (29), the parameters can be written as

$\tau_E * F_M = \mathbf{P}_E \bullet \mathbf{G}_M + W_M = \mathbf{P}_M \bullet \mathbf{G}_E + W_E = (\mathbf{P}_M - \mathbf{P}_{E_M})\bullet \mathbf{G}_E$ (65),

$\tau_M * F_M = \mathbf{P}_M \bullet \mathbf{G}_M + W_M = F_M \Rightarrow \tau_M = 1$ (66).

If the midpoint $\mathbf{P}_M$ is

1. outside the conic, at the same side of the chord $\mathbf{P}_{E_C}\mathbf{P}_{E_B}$ as $\mathbf{P}_E$ then $b_{F_M} = \overline{b_{DET}}$ and
   $b_{(\mathbf{P}_M-\mathbf{P}_{E_M})\bullet\mathbf{G}_E} = b_{F_E} = \overline{b_{DET}}$ , $b_{\tau_E} = \overline{b_{F_M} \oplus b_{(\mathbf{P}_M-\mathbf{P}_{E_M})\bullet\mathbf{G}_E}} \Rightarrow \tau_E > 0$ (67),
2. inside the conic, at the same side of the chord $\mathbf{P}_{E_C}\mathbf{P}_{E_B}$ as $\mathbf{P}_E$ then $b_{F_M} = b_{DET}$ and
   $b_{(\mathbf{P}_M-\mathbf{P}_{E_M})\bullet\mathbf{G}_E} = b_{F_E} = \overline{b_{DET}}$ , $b_{\tau_E} = \overline{b_{F_M} \oplus b_{(\mathbf{P}_M-\mathbf{P}_{E_M})\bullet\mathbf{G}_E}} \Rightarrow \tau_E < 0$ (68),
3. inside the conic, at the other side of the chord $\mathbf{P}_{E_C}\mathbf{P}_{E_B}$ as $\mathbf{P}_E$ then $b_{F_M} = b_{DET}$ and
   $b_{(\mathbf{P}_M-\mathbf{P}_{E_M})\bullet\mathbf{G}_E} = \overline{b_{F_E}} = b_{DET}$ , $b_{\tau_E} = \overline{b_{F_M} \oplus b_{(\mathbf{P}_M-\mathbf{P}_{E_M})\bullet\mathbf{G}_E}} \Rightarrow \tau_E > 0$ (69).

So, we have OoA if the midpoint $\mathbf{P}_M$ is inside the conic at the same side of the chord $\mathbf{P}_{E_C}\mathbf{P}_{E_B}$ as $\mathbf{P}_E$ , which is always at the outside of the conic. We call the area bounded by the chord $\mathbf{P}_{E_C}\mathbf{P}_{E_B}$ and the conic, the *OoA-segment*, which has the next properties:

- the measurement is inaccurate when the midpoint is inside or on the OoA-segment,
- applying the arithmetic mean equation to the points $\mathbf{P}_{E_C}$ and $\mathbf{P}_{E_B}$ gives

$$\lambda_{E_M} = \frac{\left(\mathbf{P}_{E_C} - \mathbf{P}_{E_B}\right)\bullet\left(\mathbf{G}_{E_C} - \mathbf{G}_{E_B}\right)}{4} = -F_{E_M} \quad (70),$$

- $|F_M| < |\lambda_{E_M}|$ if $\mathbf{P}_M$ is on or inside the OoA-segment (71),
- $b_{F_M+\lambda_{E_M}} = \overline{b_{F_M}}$ if $\mathbf{P}_M$ is inside the OoA-segment (72).

Hence, for $\tau_E F_M = \tau_{E_M}(F_M + \lambda_{E_M})$ , we have for (72) $\dfrac{\tau_{E_M}}{|\tau_{E_M}|} = -\dfrac{\tau_E}{|\tau_E|}$ . Therefore, if we use $F_M + \lambda_{E_M}$ instead of $F_M$ , the inaccuracy disappears. Unluckily, the exact value of $\lambda_{E_M}$ is unknown. If we replace $\lambda_{E_M}$ with $\lambda_M$ (two-point method), we enlarge the OoA-segment . This means that we will have many false OoA-events.

## 8. COMPARISON WITH PREVIOUS WORK

The midpoint methods use about the same measurements, and the general midpoint algorithms [3], [16], [4], [5, pp. 947, 951-961], [6] also use measurements to detect quadrant change control instead of using the monotonic approach. The monotonic approach is better:

- it allows the deterministic, time-invariant modeling;
- the frame can be determined beforehand;
- no quadrant change control problems;
- the general algorithms translate the curve mostly to (0,0), but the frame can only be determined after several tryouts.

Therefore, we will only compare the Berserkless Midpoint Algorithm with other midpoint algorithms with a monotonic approach.

**1. Comparison with algorithm T of D. Knuth [1, pp. 44-48, Exercise 182 at pp.47, 66 and 179]**

This is a 4-connected midpoint algorithm, not based on [7], [5]. It is ultra fast, but it is not 100% stable. If you delete the results of the OoC-Rule and if you set $b_H = b_V = 0$, then you obtain, about the same non-stable algorithm, after some simplifications. The proposed corrections [1, pp. 183] are comparable with others: they consider the next successive quadrant(s), and therefore the next monotonic direction(s). Sometimes it works, but essential it is wrong, because the monotonic direction is given a priory, and looking around the corner, neglects what is wrong in the actual monotonic segment.

**2. Comparison with Van Aken & Novak [7]**

They use separate algorithms for separate conics, and their separate conics are monotonic. They state that the accuracy $\leq \frac{\text{tolerance range}}{2}$, and that from empirical measurements, the midpoint method is better than the two-point-method; we have done the same experiments and obtained the same result, for grosso modo all the examples of [3] including thin and sharp conics.

Van Aken & Novak state, at page 166-168, the third OoC-condition: $\left|\frac{4\lambda_H}{F_D - F_B}\right| < 1$, with $4\lambda_H \triangleq (\mathbf{P}_D - \mathbf{P}_B)\bullet(\mathbf{G}_D - \mathbf{G}_B) = B\Delta^2$, $F_D - F_B = -S_{Lxy}|Y_H|\Delta$.

For the M-measurement, the third OoC-condition becomes $\left|\frac{4\lambda_M}{F_C - F_B}\right| < 1$ (73).

This can be proved very easily, if the primary condition, and therefore the 2nd OoC-condition is true.

**Proof:** $\mathbf{G}_C = \mathbf{G}_M + \frac{\mathbf{G}_C - \mathbf{G}_B}{2}$, and $\mathbf{G}_B = \mathbf{G}_M - \frac{\mathbf{G}_C - \mathbf{G}_B}{2}$. Dot multiplying with $(\mathbf{P}_C - \mathbf{P}_B)\bullet$ gives respectively $(\mathbf{P}_C - \mathbf{P}_B)\bullet\mathbf{G}_C = (\mathbf{P}_C - \mathbf{P}_B)\bullet\mathbf{G}_M + \frac{(\mathbf{P}_C - \mathbf{P}_B)\bullet(\mathbf{G}_C - \mathbf{G}_B)}{2}$ and

$$(\mathbf{P}_C - \mathbf{P}_B)\bullet\mathbf{G}_B = (\mathbf{P}_C - \mathbf{P}_B)\bullet\mathbf{G}_M - \frac{(\mathbf{P}_C - \mathbf{P}_B)\bullet(\mathbf{G}_C - \mathbf{G}_B)}{2}.$$

From the incremental equation (23), and the arithmetic mean equation (21), we have,

$$2(\mathbf{P}_C - \mathbf{P}_B)\bullet\mathbf{G}_C = (F_C - F_B)*(1 + \frac{4\lambda_M}{F_C - F_B}), \text{ and } 2(\mathbf{P}_C - \mathbf{P}_B)\bullet\mathbf{G}_B = (F_C - F_B)*(1 - \frac{4\lambda_M}{F_C - F_B}).$$

From the primary OoC-conditions (38) $b_{(\mathbf{P}_C - \mathbf{P}_B)\bullet\mathbf{G}_C} = b_{(\mathbf{P}_C - \mathbf{P}_B)\bullet\mathbf{G}_B} = b_{(F_C - F_B)}$, therefore $\left|\frac{4\lambda_M}{F_C - F_B}\right| < 1$.

Van Aken & Novak do not state the primary and secondary OoC-conditions, and we do not use directly the 3th OoC-condition. Their statement to avoid OoC, based on the 3th OoC-condition, is as acknowledged by them, completely wrong. *The third OoC-conditions and the secondary OoC-conditions are the result of the primary conditions*; they are not wrong, but these conditions do not give a 100% stable midpoint algorithm.
Van Aken & Novak [7, pp. 168] agree that their algorithms are not successful in a region in which two edges of a curve meet or cross each other. D. Knuth [1, pp. 46] avoids this situation by its a priori condition : "No edge of the integer grid contains two roots of Q".
When we apply the OoC-Rule:

- conics which have two roots on the line segment, connecting the candidate points, are not excluded;
- sharp, very sharp turning and thin conics are not excluded;
- unusual cases do not tend to drive the conics berserk.

**3. comparison with all midpoint algorithms**

All general, non-line midpoint algorithms are not 100% stable, therefore they cannot be transformed to hardware. All non-line midpoint methods, except [8], cannot be linked, directly, with the "relative curve distance" that measures $\boldsymbol{\rho}_C^2 - \boldsymbol{\rho}_B^2$.
All earlier non-line midpoint algorithms do not apply the primary conditions, nor the the OoC-rule; and most of them even do not apply the 2[nd] conditions.

## 9. CONCLUSION

Before, the digitization of conics, using the midpoint or two-point method, had a bad reputation:

- "Algorithms for discrete geometry are notoriously delicate: unusual cases tend to drive them berserk", and "Reasonable conics don't make such sharp turns" [1, pp. 180];
- "The generation of thin and sharp turning hyperbolas remains unsolved" [3, pp. 36];
- "Digitizing general conics is very hard, the octant-changing test is tougher, the difference computations are tougher. Do it only if you have to" [6, pp. 42, course cs123].

The phrases "midpoint technique" and "midpoint method" must be replaced by "midpoint measurement". The midpoint measurement finds the shortest distance from two points to a conic or QSIC, provided that the starting point is optimal, and the measurement is valid and not OoA; but even in the latter case, the tolerance of the measurement is $\leq \frac{\sqrt{2}}{2}\Delta$ in 2D and $\leq \frac{\sqrt{3}}{2}\Delta$ in 3D. Every digitized curve has non-zero tolerance and the OoA' s of the midpoint measurements are unimportant when they do not influence the tolerated tolerance.

The midpoint algorithm is ultra fast using the valid measurements or the OoC-rule, it is robust and 100% stable using the OoC-rule and it is appropriate to be converted in hardware.

We also solved the long existing enigma between the midpoint method [5, 6, 7] and the two-point method [8], and even the mystery of the midpoint method.
The Berserkless Midpoint Algorithm can be extended to 3D-QSIC-curves (intersections of quadrics).

CNC-machines need the grid points, and this poses a real problem for QSICS. Mathematica cannot calculate the QSIC itself, but it only shows the QSIC. The most difficult task is the calculation of the extreme rational tangent points.

## 10. APPENDICES

**Appendix 1** : Proof of the identity of section 5

Lagrange's identity for the vectors $\mathbf{A}$ and $\mathbf{B}$ is $\mathbf{A}^2 * \mathbf{B}^2 = (\mathbf{A} \bullet \mathbf{B})^2 + (\mathbf{A} \times \mathbf{B})^2$.

Hence, with $\mathbf{A} = \boldsymbol{\rho}_i$ and $\mathbf{B} = \mathbf{G}_E$, we have $(\boldsymbol{\rho}_i \times \mathbf{G}_E)^2 = \boldsymbol{\rho}_i^2 * \mathbf{G}_E^2 - (\boldsymbol{\rho}_i \bullet \mathbf{G}_E)^2$.

From (15), we have $\mathbf{G}_E = S_{LEFT}(\mathbf{T}_{p_E} \times \mathbf{k})$, and applying the vector triple product

$\mathbf{A} \times (\mathbf{C} \times \mathbf{D}) = (\mathbf{A} \bullet \mathbf{D}) * \mathbf{C} - (\mathbf{A} \bullet \mathbf{C}) * \mathbf{D}$ with $\mathbf{C} = \mathbf{T}_{p_E}$ and $\mathbf{D} = \mathbf{k}$ gives

$\boldsymbol{\rho}_i \times (\mathbf{T}_{p_E} \times \mathbf{k}) = (\boldsymbol{\rho}_i \bullet \mathbf{k}) * \mathbf{T}_{p_E} - (\boldsymbol{\rho}_i \bullet \mathbf{T}_{p_E}) * \mathbf{k} = -(\boldsymbol{\rho}_i \bullet \mathbf{T}_{p_E}) * \mathbf{k}$ because the vector $\mathbf{k}$ is perpendicular to the plane containing the distance vector $\boldsymbol{\rho}_i$, the tangent $\mathbf{T}_{p_E}$, and the gradient $\mathbf{G}_E$.

Hence $(\boldsymbol{\rho}_i \times \mathbf{G}_E)^2 = \left(-S_{LEFT}(\boldsymbol{\rho}_i \bullet \mathbf{T}_{p_E}) * \mathbf{k}\right)^2 = (\boldsymbol{\rho}_i \bullet \mathbf{T}_{p_E})^2 = \boldsymbol{\rho}_i^2 * \mathbf{G}_E^2 - (\boldsymbol{\rho}_i \bullet \mathbf{G}_E)^2$

**Appendix 2** : Average %-speed gain versus Mathematica's ContourPlot[]

| Berserkless Midpoint Algorithm | Mathematica version 10.00 | $\left\lvert\frac{\text{Btime} - \text{Mtime}}{\text{Mtime}}\right\rvert$ |
|---|---|---|
| Btime | Mtime | |
| 8.122465 | 11.85668 | 31.49 % |
| 8.007458 | 10.15458 | 21.14 % |
| 7.920453 | 10.77762 | 26.51 % |
| 7.904452 | 10.90962 | 27.55 % |
| Average %-speed gain using the Berserkless Midpoint Algorithm | | 26.67 % ± 5% |

Mathematica's AbsoluteTime[] measures the execution time of both algorithms. The generation of the drawings is the only difference of the two algorithms.

**Appendix 3**: Examples of bad and good digitalizations

Ellipses (Algorithm T of D. Knuth digitizes B-4 and D-4) :

| | A | B | C | D |
|---|---|---|---|---|
| | F(x,y)=x² + 15² y² - 15²<br>From (0,-1) to (15,0) | | F(x,y)=-160x²-921y²+767xy-104x+249y | |
| 8 | 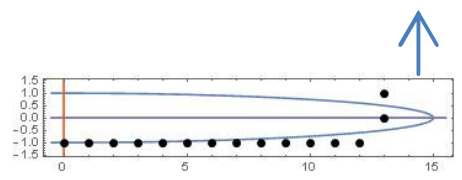 | 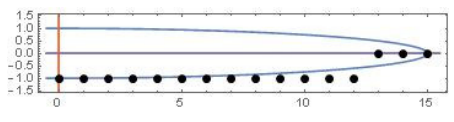<br>OoC-rule / Finish horizontally | 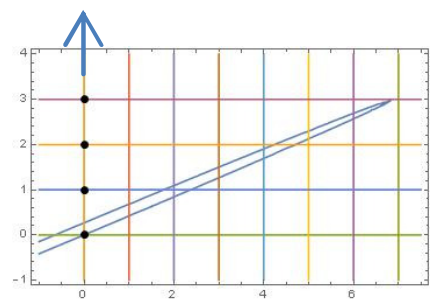 | 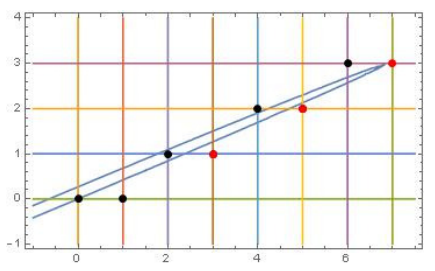 |
| 4 | 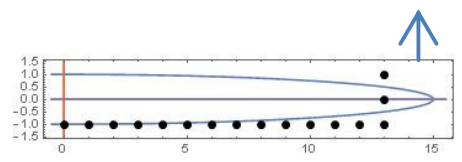 | Alg. T of D. Knuth<br>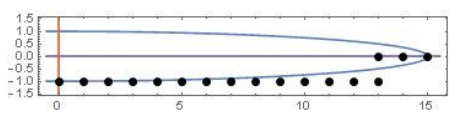<br>Finish horizontally | From (0,0) to (7,3)<br>OoC-rule digitizes the red points | 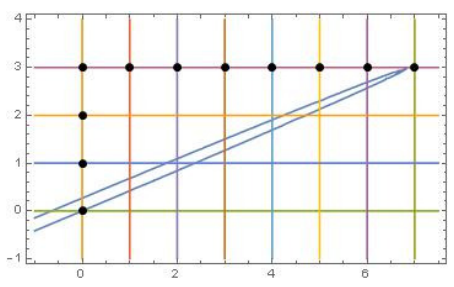 |
| | Non-monotonic & Berserk | Monotonic algorithm | Non-monotonic & Berserk | Berserkless midpoint algorithm D-8 |

<table>
<tr><td colspan="3">Wide hyperbola $F[x,y]=24x^2+4y^2+2*10xy+2*17x+2*7y+8$, but figures rotated $90°$ CCW</td></tr>
<tr><td>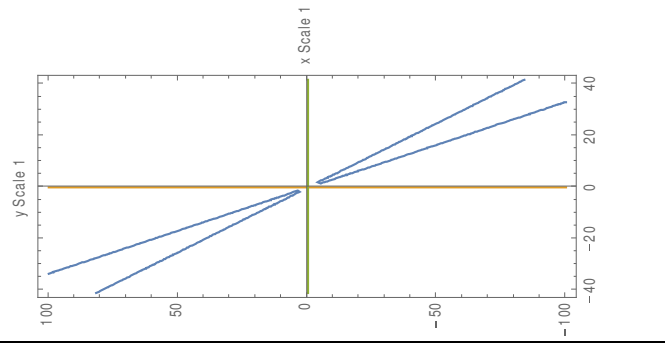</td><td>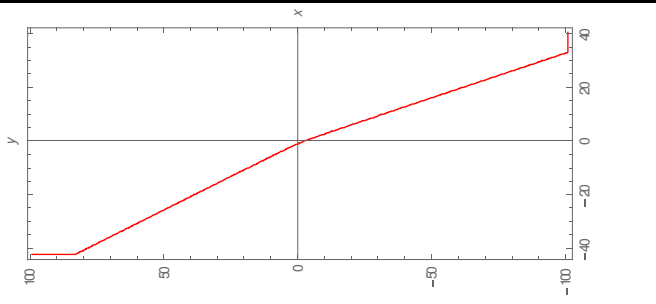</td><td>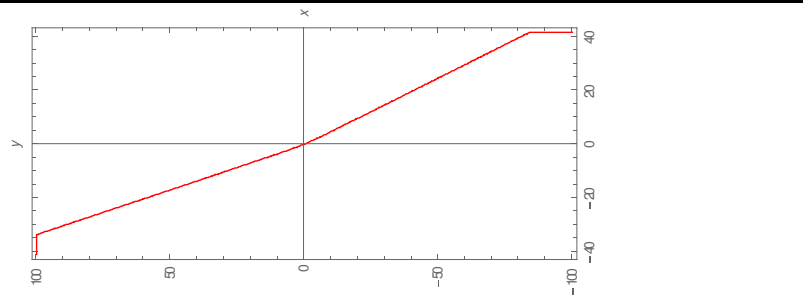</td></tr>
<tr><td>From Mathematica, it suggests a hyperbola with two thin branches</td><td colspan="2">Wide branches of hyperbola, generated by the berserkless midpoint algorithm. The line y+0.5=-2.5(x+0.5) does not cut the hyperbola.</td></tr>
</table>

**Appendix 4** Simple example which shows that Algorithm T of D. Knuth [1] can be OoA:

T-algorithm of D. Knuth using the notations and formulas of [1, pp. 46-47]:

$F[x,y]=20x^2+20y^2-291$, $Q[x,y]=F[x-0.5,y-0.5]=20x^2+20y^2-20x-20y-281$, $a=c=20, b=0, d=c=-20, f=-281$.

Cases T2,T3,T4,T5 of D. Knuth correspond with the direction vectors T1,T4,T2,T3.

Initialization T1:

Q[4,0]=-41, Qx[4,0]=140, Qy[4,0]=-20, QM=Q=Q[x,y+1]=-41, RM=R=Qx[x,y+1]=120, SM=S=Qy[x,y+1]=40 (see [1, pp 45 & 46 and (179))

*Loop* T4:

| x | y | QM or Q | RM or R | SM or S | This is loop T4 |
|---|---|---|---|---|---|
| 4 | 0 | -41 | 120 | 40 | Do T6 if QM < 0 |
| 4 | 1 | -1 | 120 | 80 | Point (4,1) is Out-of-Accuracy |
| 4 | 2 | 79 | 120 | 120 | Point (3,1) is better |
| 3 | 2 | -41 | 80 | 120 | |
| 3 | 3 | 79 | 80 | 160 | Do T8 if QM >=0 |

FR=ImplicitRegion[F[x,y]==0,{x,y}], d[xi_,yi_]=N[RegionDistance[FR,{xi,yi}]
d[4,2]=0.658 > d[3,1]=0.652 for $R^2 = 291 / 20$ and
d[4,2]=0.664 > d[3,1]=0.646 for $R^2 = 290 / 20$ (midpoint on the circle).

Note that D. Knuth's QM function is identical with the midpoint function $F_M$.

## 11. ACKNOWLEDGMENT

We wish to acknowledge the careful review and helpful comments of Dr. H. V. Ramakrishnan, Dr MGR Educational & Research Institute University Maduravoyal, Chennai.